\documentclass[twocolumn,showpacs,preprintnumbers,amsmath,amssymb]{revtex4}
\usepackage{graphicx}% Include figure files
\usepackage{dcolumn}% Align table columns on decimal point
\usepackage{bm}% bold math

\begin{document}

%\preprint{APS/123-QED}

\title{Spectroscopic Signatures on Increase in Charge-Density-Wave Potential of 1{\em T}-TaS$_{1.2}$Se$_{0.8}$}

\author{Y. Aiura}
 \email{y.aiura@aist.go.jp}
\author{I. Hase}
\author{H. Bando}
\author{K. Yagi-Watanabe}
\affiliation{
National Institute of Advanced Industrial Science and Technology, Tsukuba, Ibaraki 305-8568, Japan}

\author{K. Tanaka}
\affiliation{
Department of Complexity Science and Engineering, University of Tokyo, Tokyo 113-0033, Japan}

\author{K. Ozawa}
\affiliation{
Department of Chemistry and Materials Science, Tokyo Institute of Technology, Tokyo 152-0033, Japan}

\author{T. Iwase}
\author{Y. Nishihara}
\affiliation{
Faculty of Science, Ibaraki University, Mito, Ibaraki 310-8512, Japan}

\author{O. Shiino}
 \altaffiliation[Present address:]{
R $\&$ D Division, Bridgestone Corporation, Kodaira, Tokyo 187-8531, Japan}

\author{M. Oshima}
\affiliation{
Department of Applied Chemistry, University of Tokyo, Tokyo 113-8656, Japan}

\author{M. Kubota}
\author{K. Ono}
\affiliation{
Institute of Materials Structure Science, Tsukuba, Ibaraki 305-0801, Japan}

\date{\today}% It is always \today, today,

\begin{abstract}
We present an angle-resolved photoemission (ARPES) study on the layered transition-metal dichalcogenide 1{\em T}-TaS$_{1.2}$Se$_{0.8}$ in the metallic commensurate charge-density-wave (CDW) phase.  A model calculation of the spectral function captures the main features of the ARPES spectra well qualitatively, that is, the gross splits of unreconstructed band structure in the absence of the CDW superlattice.  The observed enhancement of the size of the gap between the lower and middle fragments of the Ta {\em 5d} band along the $\Gamma$M line by cooling is interpreted in terms of the increase in the CDW-related potential.
\end{abstract}

%**********************************
\maketitle

% ********** INTRODUCTION **********
Quasi-two-dimensional (quasi-2D) materials, like the layered transition metal dichalcogenides, 1{\em T}-TaS$_{x}$Se$_{2-x}$, have attracted much attention because of their various unique physical properties combined with the formation of charge-density wave (CDW) \cite{Wilson75}.  1{\em T}-TaS$_{2}$ shows a metal-to-insulator (MI) transition at 180 K which occurs followed by nearly commensurate (NC) to commensurate (C) CDW transition \cite{Kim94a}.  The MI transition is now understood in terms of a Mott localization, triggered by the NC-C transition \cite{Fazekas80}.  On the other hand, isostructural and isoelectronic 1{\em T}-TaSe$_{2}$ remains metallic to very low temperature, suggesting that the Ta 5d electrons in Se compound are less susceptible to the Mott localization than those in S one \cite{Perfetti03}.  The MI transition of 1{\em T}-TaS$_{x}$Se$_{2-x}$ occurs between x=1.2 and x=1.5 \cite{Shiino96,Shiino00,Horiba02a}.  Spectroscopic signatures of an energy gap at the Fermi energy (E$_{F}$) were observed in the C phase in the sample with x=1.5 by scanning tunneling spectroscopy (STS), but not in the sample with x=1.2 \cite{Shiino96}.  1{\em T}-TaS$_{1.2}$Se$_{0.8}$ exhibits only one (metallic) C phase below the incommensurate to C transition temperature, but the resistivity in the C phase is complicated \cite{Shiino96,Horiba02a}.  Therefore, 1{\em T}-TaS$_{1.2}$Se$_{0.8}$ is suitable system for studying the anomalous metallic behavior near the Mott localization,  because it is possible to control the potential of the superlattice {\em in the same CDW phase}.  

Angle-resolved photoemission spectroscopy (ARPES) and STS, which probes the 
single-particle spectral function, can provide a direct view of changes in the 
dramatic rearrangement of the electronic structure near the MI transition.  
The importance of the electron correlation effects in 1{\em T}-TaS$_{2}$ and 
1{\em T}-TaSe$_{2}$ has been recognized enough by virtue of various STS 
\cite{Kim94a,Kim94b} and ARPES \cite{Smith85,Manzke89,Claessen90,Dardel92,Zwick98,Pillo99,Pillo00,Horiba02b,Aiura03a,Perfetti03}.   In this paper, we report ARPES results on 1{\em T}-TaS$_{1.2}$Se$_{0.8}$ in the metallic C phase taken at 300 K and 90 K.  It was shown that the size of the gap, which was observed between the lower and middle fragments of the Ta {\em 5d} band at the 0.65$\Gamma$M point, grows by cooling.  From model calculations, it can be understood that the gap size is closely connected with the CDW-related potential.  The width of the peak near E$_{F}$ in the calculated density of states (DOS) narrows with the increase in the CDW-related potential, which lends support to the notation of the anomalous metallic behavior near the Mott transition.

% ********** CALCULATION **********
In previous empirical tight-binding calculation of 1{\em T}-TaS$_{2}$ in the presence of commensurate CDW superlattice by Smith {\em et al.}, it has been shown that the Ta-derived {\em d} band collapses into three sub-manifolds separated by gaps \cite{Smith85}.  Signatures of those manifolds and/or gaps were partially confirmed by many ARPES experiments \cite{Smith85,Claessen90,Manzke89,Dardel92,Zwick98}.  However, it remains to be explained why the experimental band dispersive behavior of those manifolds except those gaps resemble so closely the theoretical band calculations in the normal state in the absence of the CDW superlattice, and why the three bands in the lower and middle manifolds could not be observed clearly even using the high-resolution ARPES \cite{Zwick98}.  From a recent ARPES study, moreover, it was shown that the remnant Fermi surface of 1T-TaS$_{2}$ exhibits the symmetry of the one-particle normal state one \cite{Pillo99}.  Those ARPES results mean that the overall shape of the electron wave functions is governed by the Fourier components of the crystal potential for the unreconstructed lattice, rather than the CDW-related contribution \cite{Claessen90,Aiura03a}.  In such weak competing potentials, recently, Voit {\em et al.} proposed model calculations on a quasi-one-dimensional (quasi-1D) material \cite{Voit00}.  The band structure and spectral weight distribution obtained by diagonalizing the Hamiltonian truncated {\em at the first order} reproduced ARPES spectra very well.  To interpret the ARPES spectra of quasi-2D material 1{\em T}-TaS$_{x}$Se$_{2-x}$ we developed the model calculations proposed for a quasi-1D material.

First, we calculated band structure in absence of CDW superlattice using empirical tight-binding calculations \cite{Smith85}.  Since the d$_{yz}$ and d$_{zx}$ orbitals reside at energies well above E$_{F}$, we can drop them from the basis set.  Therefore, three bands that consist of the d$_{3z^{2}-r^{2}}$, d$_{x^{2}-y^{2}}$ and d$_{xy}$ orbitals are shown in the calculation.  Since only the lowest band among them crosses E$_{F}$ and is occupied, we treated it mainly in the model calculations.  The Hamiltonian is
\begin{equation}
 H_{{\bf k+g_{i}},{\bf k+g_{j}}}=\varepsilon({\bf k}+{\bf g_{i}})\delta_{{\bf g_{i}},{\bf g_{j}}}+[V_{{\bf g_{i}},{\bf g_{j}}}c^{\dag}_{{\bf k+g_{i}}}c_{{\bf k+g_{j}}}+H.c.].
\end{equation}
$\varepsilon({\bf k+g_{j}})$ is the kinetic energy of the lowest band, the potential is $V_{{\bf g_{i}},{\bf g_{j}}}=V\delta_{|{\bf g_{i}}-{\bf g_{j}}|,g}$, ${\bf g_{i}}$ is one of the reciprocal lattice points of the $\sqrt{13}\times\sqrt{13}$ superstructure in the unreconstructed $1\times1$ BZ ($i$=0, . . ., 12), and {\em g} describes the size of the smallest reciprocal lattice vector ($\frac{4}{\sqrt{13}}\frac{\pi}{a}$, where $a$ is lattice constant).  The single-particle spectral function $\rho({\bf k},\varepsilon)$ is given by
\begin{equation}
\rho({\bf k}+{\bf g_{i}},\varepsilon)=\sum_{j=0}^{12} [a_{j}({\bf k})]_{i}^{2}\delta[\varepsilon-E_{j}({\bf k})],
\end{equation}
in terms of the components of the eigenvectors $a_{j}$ associated with eigenvalues $E_{j}$ of the Hamiltonian (1).

\begin{figure}
	\includegraphics{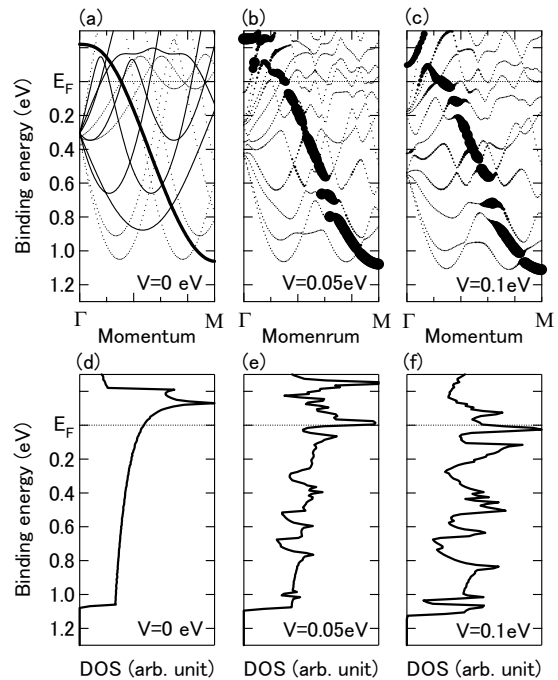}
	\caption{(a) Unreconstructed band of 1{\em T}-TaS$_{1.2}$Se$_{0.8}$ along the $\Gamma$M line without the CDW superlattice (thick curve) and the Umklapp bands (thin and dotted curves) in the case for V=0 eV.  The single-particle spectral function in the case for (b) V=0.05 and (c) 0.10 eV.  The size of the symbols is proportional to the spectral weight.  Calculated Ta {\em d} band density of states in the case for (d) V=0, (e) V=0.05, and (f) V=0.10 eV, respectively.}
\end{figure}

In the case for V=0 eV, the band of 1{\em T}-TaS$_{1.2}$Se$_{0.8}$ along the $\Gamma$M line in absence of the CDW superlattice (thick curve) is shown in Fig. 1 (a).  No spectral weight is expected for any of the additional bands derived from the Umklapp process (thin and dotted curves), i.e., weak Bragg diffraction through reciprocal-lattice vectors of the CDW superlattice.  Figure 1 (b) and (c) show the single-particle spectral function in the case for V=0.05 and 0.10 eV, respectively.  Figures 1 (d), (e), and (f) shows the calculated DOS for V=0, 0.05 and 0.10 eV, respectively.  To compare with the ARPES spectra, the single-particle spectral function was multiplied by the Fermi-Dirac distribution function at 300K or 90K, and then convoluted with Gaussian functions of 150 meV and $\pm0.1\AA^{-1}$ FWHM which represent the instrumental energy and momentum resolution, respectively.  Figure 2 (a) and (c) show the intensity plot and the energy distribution curves (EDCs) from the calculated single-particle spectral weight function in the case for V=0.05 eV and T=300K.  Figure 2 (b) and (d) show the calculated intensity plot and the EDCs in the case for V=0.10 eV and T=90K.

\begin{figure}
	\includegraphics{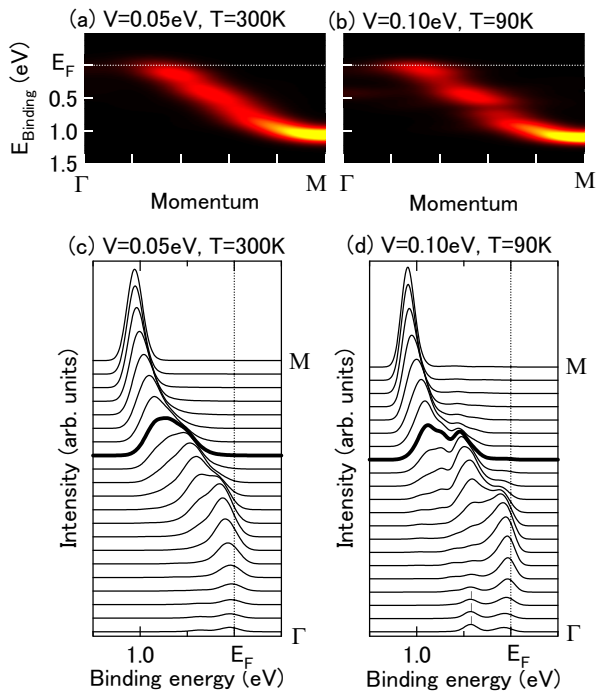}
	\caption{(Color online) (a) Intensity plot and (c) EDCs from the calculated single-particle spectral weight function in the case for V=0.05 eV and T=300K.  (b) The intensity plot and (d) the EDCs in the case for V=0.10 eV and T=90K.  Bright and dark parts correspond to high and low spectral weight, respectively.}
\end{figure}

% ********** EXPERIMENTAL **********

We performed ARPES measurements at BL-1C of the Photon Factory (KEK, Tsukuba) using an electron spectrometer mounted on a two-axis goniometer (VG ARUPS10) \cite{Ono01}.  The sample goniometer used here provides independent polar, azimuth and tilt rotation of the sample (R-Dec Co. Ltd., {\em i} GONIO LT) \cite{Aiura03b}.  All ARPES spectra were taken at the photon energy (h$\nu$) of 40 eV.  The samples were mounted vertically and only photoelectrons emitted from the plane defined by the light beam and the surface normal were observed.  The emission angle of the photoelectron measured from the surface normal was varied by rotating the energy analyzer horizontally, whereas the angle of incidence of the light was fixed to $45^{\circ}$.  The azimuth angle was varied by rotating the samples to the surface normal.  Single crystals of 1{\em T}-TaS$_{1.2}$Se$_{0.8}$ were grown by the iodine transport technique and characterized by resistivity measurement \cite{Shiino96,Horiba02a}.  The samples were cleaved {\em in situ} at a base pressure of $3\times10^{-10}$ Torr.  We measured ARPES spectra at 300 K and at 90 K.  The energy and spatial resolution were 0.15 eV and $\pm2^{\circ}$, respectively.

% ********** RESULTS AND DISCUSSION **********

\begin{figure}
	\includegraphics{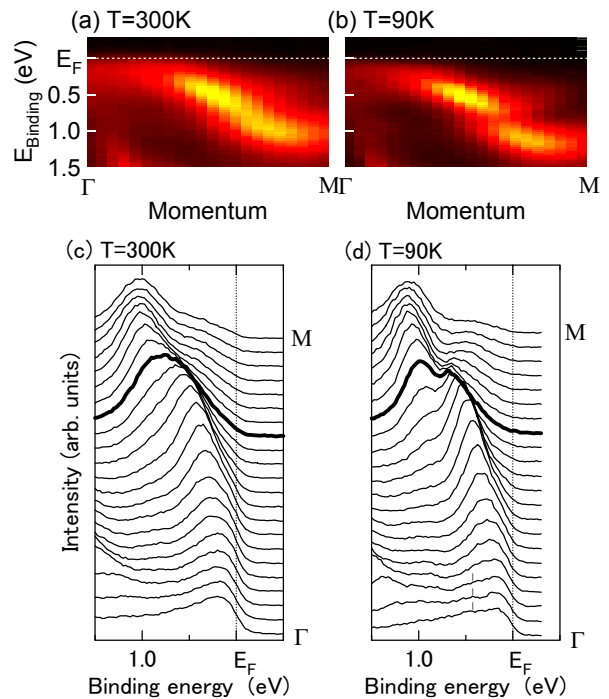}
	\caption{(Color online) ARPES intensity plots at (a) 300K and (b) 90K and EDCs at (c) 300 K and (d) 90K along the high symmetry $\Gamma$M line (h$\nu$=40eV).}
\end{figure}

Figure 3 shows ARPES intensity plots at (a) 300K and (b) 90K and EDCs at (c) 300K and (d) 90K along the high symmetry $\Gamma$M line.  At first sight, it is shown in Fig. 3 (a) that a Ta {\em 5d} band disperses downward in energy along the high symmetry line from the $\Gamma$ point to the M point.  Contrary to the prediction of previous band calculations of 1{\em T}-TaS$_{2}$ \cite{Mattheiss73,Myron75,Woolley77} and 1{\em T}-TaSe$_{2}$ \cite{Myron75,Woolley77}, however, the observed Ta {\em 5d} band never crosses E$_{F}$ and lie around the $\Gamma$ point at a binding energy of about 0.2 eV, which is consistent with the previous ARPES spectra in 1{\em T}-TaS$_{2}$ in the metallic NC phase \cite{Pillo99,Pillo00} and 1{\em T}-TaSe$_{2}$ in the metallic C phase \cite{Aiura03a}.  Furthermore, the peak width of the Ta {\em 5d} band is broadened at the 0.65$\Gamma$M point (thick curve in Fig. 3 (c)).  This broad peak is divided into two structures by cooling (thick curve in Fig. 3 (d)).  As the result, the intensity plot at 90K in Fig. 3 (b) shows the split of the Ta {\em 5d} band at the 0.65$\Gamma$M point clearly.  In addition, a weak hump appears at a binding energy of about 0.4eV around the $\Gamma$ point, as shown by ticks around the $\Gamma$ point in Fig. 3 (d).

It is obvious to reproduce those spectral features of the Ta {\em 5d} band by the model calculations well, except the absence of crossing over E$_{F}$.  The split of the Ta {\em 5d} band at the 0.65$\Gamma$ point and the appearance of the weak hump at a binding energy of about 0.4eV around the $\Gamma$ point is ascribed to the increase in V as shown in Fig. 2.  Judging from comparison of the ARPES spectra at the 0.65$\Gamma$M point in Figs. 3 (c) and (d) (thick curves) and the corresponding spectral weight distributions in Figs. 2 (c) and (d) (thick curves), the potentials for the ARPES spectra at 300 K and 90 K were estimated at 0.05 eV and 0.1 eV, respectively.  Although the split of the Ta {\em 5d} band in 1{\em T}-TaS$_{2}$ and 1{\em T}-TaS$_{1.5}$Se$_{0.5}$ in the metallic NC phase is almost equal to that in 1{\em T}-TaS$_{1.2}$Se$_{0.8}$ shown here, it is dramatically enhanced in the insulating C phase, meaning that the influence of the CDW-related potential became dramatically large in the insulating C phase \cite{Zwick98,Horiba03}.  With increasing V, the calculated spectral weight distribution shifts to a low binding energy side (not shown), which is consistent with the observed shift of the ARPES spectra in 1{\em T}-TaS$_{1.5}$Se$_{0.5}$ due to the NC-C phase transition \cite{Horiba02a,Horiba03}.

The calculated intensity plot and the EDCs for V=0.10 eV show another gap near E$_{F}$ at about 0.4$\Gamma$M due to the split of the Ta {\em 5d} bandas as shown in Figs. 2(b) and (d).  That is, the Ta {\em 5d} band splits into three fragments under the influence of the CDW-related potential.  The gap near E$_{F}$ at the 0.4$\Gamma$M point was well observed for 1{\em T}-TaS$_{2}$ in the insulating C phase \cite{Zwick98}.  Also for 1{\em T}-TaS$_{2}$ in the metallic NC phase \cite{Pillo00} and for 1{\em T}-TaSe$_{2}$ in the metallic C phase \cite{Aiura03a}, the gap was not clear because the upper fragment in the metallic phase is very broad compared with that in the insulating phase.  No crossing of the Fermi level is supposed to be induced by the electron correlation effects in the Ta {\em 5d} band.  According to the model of Tosatti and Fazekas \cite{Fazekas80}, the (partially filled) upper fragment straddling E$_{F}$ split into the lower Hubbard band (LHB) and the upper Hubbard band (UHB) by the electron correlation effects.  For 1{\em T}-TaS$_{2}$ in the insulating C phase the flat LHB was observed around the $\Gamma$ at a binding energy of 0.19 eV \cite{Zwick98} and in the metallic NC phase \cite{Pillo00}, and at about 0.2 eV for 1{\em T}-TaSe$_{2}$ in the metallic C phase \cite{Aiura03a}.  For the metallic phase, E$_{F}$ lies in a pseudogap created by the tails of two overlapping Hubbard subbands.  Based on our model calculation, the width of the upper fragment straddling E$_{F}$, i.e., the width of the Hubbard bands, depends on the strength of the potential, V.  This is supported by the calculated DOS spectra in Figs. 1 (d), (e) and (f).  No peak in DOS for V=0 eV is shown near E$_{F}$ in Fig. 1(d).  A broad peak appears under V=0.05 eV (Fig. 1 (e)) and becomes sharp with increasing the strength of V to 0.1 eV (Fig. 1 (f)).

\begin{figure}
	\includegraphics{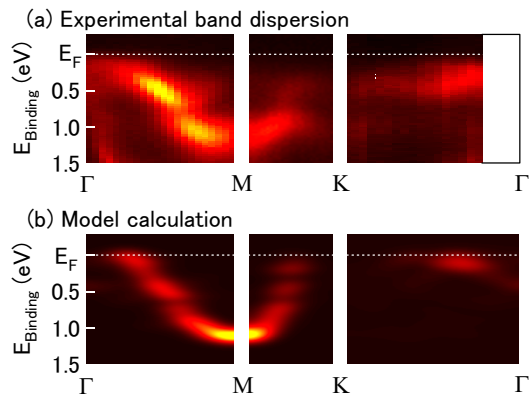}
	\caption{(Color online) Experimental band dispersion at 90 K along the high symmetry lines $\Gamma$M, MK and K$\Gamma$.  (b) Corresponding intensity plot of the calculated spectral weight distribution in the case for V=0.1eV and T=90 K.}
\end{figure}

Figure 4 shows (a) the experimental band dispersion at 90 K along the high symmetry lines $\Gamma$M, MK and K$\Gamma$ and (b) the corresponding intensity plots of the calculated spectral weight distribution in the case for V=0.1eV and T=90 K.  It is shown that the model calculations reproduce the main features of the ARPES results very well, e.g., the dominant spectral feature from the lower subband along the MK line and the appearance of the weak spectral weight around the middle of the $\Gamma$K like.  This means that the model calculations presented here, which are originally proposed for a quasi-1D material \cite{Voit00}, are also useful for a quasi-2D material, 1{\em T}-TaS$_{x}$Se$_{2-x}$.  It is reasonable to suppose that the slight difference in the energy position of the spectral features near E$_{F}$ is based on the electron correlation effects, which are not included in the model calculation.

Since V in Eq. (1) represents the coupling strength between the unreconstructed band and the CDW-derived bands, the spectral changes by temperature may depend on the size of the atomic displacement and/or the domain size of the CDW superlattice.  To investigate the electronic behavior due to the MI transition by a Mott transition, 1{\em T}-TaS$_{2}$ is not suitable because the MI transition is triggered by the structural NC-C phase transition.  Recently, Perfetti {\em et al.} reported that a MI transition without accompanying such structural phase transition occurs at the surface of 1{\em T}-TaSe$_{2}$, which is an ideal material to tune the crucial parameter (W/U), where U is the on-site Coulomb correlation energy and W is the bandwidth \cite{Perfetti03}.  However, it is hard to know the role of the CDW-related potential in the electronic behavior of the Ta 5d in a situation close to the Mott transition since the ARPES spectra were shown only in the limited area of the Brillouin zone.  To study this, the elucidation of the entire electronic structure of the Ta {\em 5d} band is desired strongly.

% ********** SUMMARY **********
In summary, we presented ARPES spectra of 1{\em T}-TaS$_{1.2}$Se$_{0.8}$ in the metallic phase taken at 300 K and 90 K and compared them with the calculated spectra based on the model proposed by Voit {\em et al.} \cite{Voit00}.  It was shown that the calculated single-particle spectral function capture the main features of the ARPES spectra well qualitatively.  We can be fairly certain that the model calculations shown here play an important role to understand the physical properties of quasi-two-dimensional transition-metal dichalcogenides.\\

\begin{acknowledgments}
This work was partly done under Project No. 2002G174 at the Institute of Material Structure Science in KEK.
\end{acknowledgments}

%\newpage
%\bibliography{pr}% Produces the bibliography via BibTeX.

\newpage

\end{document}